# Anomalous Tunneling Magnetoresistance Oscillation and Electrically Tunable Tunneling Anisotropic Magnetoresistance in Few-layer CrPS$_4$

ZhuangEn Fu[1,2], Hong-Fei Huang[3], Piumi Samarawickrama[1,2], Kenji Watanabe[4], Takashi Taniguchi[5], Wenyong Wang[1], John Ackerman[6], Jiadong Zang[7], Jie-Xiang Yu[3*], and Jifa Tian[1,2*]

1. Department of Physics and Astronomy, University of Wyoming, Laramie, Wyoming 82071, United States
2. Center for Quantum Information Science and Engineering, University of Wyoming, Laramie, Wyoming 82071, United States
3. School of Physical Science and Technology, Soochow University, Suzhou 215006, China
4. Research Center for Electronic and Optical Materials, National Institute for Materials Science, 1-1 Namiki, Tsukuba 305-0044, Japan
5. Research Center for Materials Nanoarchitectonics, National Institute for Materials Science, 1-1 Namiki, Tsukuba 305-0044, Japan
6. Department of Chemical and Biomedical Engineering, University of Wyoming, Laramie, Wyoming 82071, United States
7. Department of Physics and Astronomy, University of New Hampshire, Durham, New Hampshire 03824, United States

*E-mail: jxyu@suda.edu.cn, jtian@uwyo.edu

ABSTRACT

Two-dimensional (2D) van der Waals (vdW) magnets with layer-dependent magnetic states and/or diverse magnetic interactions and anisotropies have attracted extensive research interest. Despite the advances, a notable challenge persists in effectively manipulating the tunneling anisotropic magnetoresistance (TAMR) of 2D vdW magnet-based magnetic tunnel junctions (MTJs). Here, we report the novel and anomalous tunneling magnetoresistance (TMR) oscillations and pioneering demonstration of bias and gate voltage controllable TAMR in 2D vdw MTJs, utilizing few-layer CrPS$_4$. This material, inherently an antiferromagnet, transitions to a canted magnetic order upon application of external magnetic fields. Through TMR measurements, we unveil the novel, layer-dependent oscillations in the tunneling resistance for few-layer CrPS$_4$ devices under both out-of-plane and in-plane magnetic fields, with a pronounced controllability via gate voltage. Intriguingly, we demonstrate that both the polarity and magnitude of TAMR in CrPS$_4$ can be effectively tuned through either a bias or gate voltage. We further elucidate the mechanism behind this electrically tunable TAMR through first-principles calculations. The implications of our findings are far-reaching, providing new insights into 2D magnetism and opening avenues for the development of innovative spintronic devices based on 2D vdW magnets.

**1. Introduction**

Since 2017, the study of two-dimensional (2D) van der Waals (vdW) magnets has become a key area of scientific and technological importance[1,2]. The 2D vdW magnets featuring distinctive properties, especially their marked layer-dependent magnetism and diverse types of magnetic interaction and anisotropy[1–25], offer ideal platforms for exploring magnetism down to 2D limit. In the past few years, different approaches/techniques have been developed to identify and manipulate magnetism and magnetic interactions of atomically thin 2D vdW magnets, including magneto-optic Kerr microscopy [1,2], single-spin microscopy[26], Raman spectroscopy[27], second-harmonic generation[28], spin-polarized scanning tunneling microscopy[11], electrostatic doping[6–8], pressure[29], circularly polarized light[30], tunneling transport[31–37], etc. Among them, an exciting development is the emergence of the 2D vdW magnet-based magnetic tunnel junctions (MTJs). Different from conventional MTJs composed of a structure of a ferromagnet (FM)/insulating spacer/FM, the 2D vdW-based MTJs utilize atomically thin 2D insulating or semiconducting magnetic layers that function both as a spin-polarized current generator and a tunneling barrier. This innovative architecture adds a new dimension to MTJs, leveraging the intrinsic characteristics of 2D magnets. Novel magnetic properties and quantum phenomena of 2D vdW magnets can be explored through tunneling measurements, harnessing the potential of next-generation spintronic devices. A notable example is graphene/$CrI_3$/graphene[31,33–37] MTJ devices. In the past few years, novel properties and device prototypes of 2D vdW-based MTJs have been explored using tunneling transport, for instance, layer-dependent 2D magnetism in $CrX_3$[31,35,38], spin tunnel field-effect transistors in $CrI_3$[33], giant magnetoresistance in $CrI_3$[34,39], magnon-assisted tunneling in $CrBr_3$[40], magneto-memristive effects in $CrI_3$[36,41], etc. While considerable advancements have been achieved in the field, the majority of research has centered on few-layer $CrI_3$-based MTJs, which are recognized for their layer-dependent magnetism, strong magnetic anisotropy, and spin-flip transition characteristics. However, considering the wide array of magnetic interactions and anisotropies, it is essential to explore a more diverse range of magnetic states/interactions, such as spin-flop transition and tunability of magnetic anisotropy in 2D semiconducting vdW magnets.

Chromium thiophosphate ($CrPS_4$)[3,42–56] represents a paradigmatic example, exemplifying the intricate interplay between its magnetic, structural, electrical, and optical properties. For instance, in monolayer CrPS$_4$, quasi-1D chains of edge-sharing CrS$_6$ octahedra extend along the *b*-axis, interconnected along the *a*-axis by PS$_4$ tetrahedra[3,42]. Bulk CrPS$_4$ exhibits intralayer ferromagnetic coupling and interlayer antiferromagnetic, classified as an A-type antiferromagnet with a Néel temperature of ~ 38 K[50]. Contrary to CrI$_3$, CrPS$_4$ not only demonstrates stability under ambient conditions[48], but also features spin canting in individual layers under an external magnetic field (**Figure 1**), leading to a spin-flop phase transition[57,60]. Furthermore, CrPS$_4$ is a magnetic semiconductor with an indirect band gap of ~ 1.4 or 1.31 eV determined by optical measurements[3,57], offering a unique opportunity for tuning their electronic and magnetic properties via gate voltages[43]. Despite growing interest in exploring the unique properties of CrPS$_4$, aspects such as its layer-dependent magnetism, quantum effects, and the potential for tunable magnetic anisotropy in its few-layer form are still largely unexplored.

In this work, based on tunneling magnetoresistance (TMR) measurement, we report layer-dependent characteristics of TMR in few-layer CrPS$_4$-based MTJs. We find that the TMR of the few-layer CrPS$_4$ shows anomalous and novle TMR oscillations with varying out-of-plane (OOP) or in-plane (IP) magnetic fields, depending on the number of layers. The TMR is further found to be highly tunable by electrostatic doping. Strikingly, we demonstrate that the polarity and magnitude of tunneling anisotropic magnetoresistance (TAMR) in few-layer CrPS$_4$ can be controlled by either a bias or a gate voltage. Lastly, we employ first-principles calculations to elucidate the underlying mechanism of the electrically tunable TAMR in few-layer CrPS$_4$. Our findings not only contribute to the fundamental understanding of 2D magnetism and novel quantum effects in vdW magnetic materials but also pave the way for developing ultra-compact, energy-efficient spintronic devices, thereby marking a significant stride in the manipulation and utilization of spin states in low-dimensional systems.

## 2. Results

### 2.1. Layer- and Gate-dependent Tunneling Magnetotransport in Few-layer CrPS$_4$

To probe the layer-dependent magnetic states in few-layer CrPS$_4$, we employed tunneling transport measurements[31,32,39]. The CrPS$_4$-based MTJs with both top and bottom gates were fabricated by the commonly used dry transfer method (see Methods). **Figure 1a** and **1b** show a schematic side view and an optical image of a bilayer (2L) CrPS$_4$-based MTJ, respectively. This device features a structure of graphite (top gate)/hBN/graphite (electrode)/CrPS$_4$/graphite (electrode)/hBN on a SiO$_2$ (285 nm)/Si substrate (back gate). The high quality of the CrPS$_4$-based tunnel junction is evidenced by the exponential increase in current with increasing bias voltage, consistently observed in the presence and absence of magnetic fields (**see Figure S1** in Supporting Informatino). It is known that bulk $CrPS_4$ is an A-type antiferromagnet and undergoes a spin-flop transition from a spin-antiparallel (SAP) state to a canted spin state at $B \approx 0.7$ T, followed by a transition to a spin-parallel (SP) state at $B \approx 8.0$ T,[50] as schematically illustrated in **Figure 1c**. **Figure 1d** shows the tunneling resistance as a function of temperature ($R$ vs. $T$) for few-layer $CrPS_4$ devices with different thicknesses, including bilayer (2L, ~ 1.2 nm), trilayer (3L, ~ 1.8 nm) and quadra-layer (4L, ~ 2.4 nm). We see that for all the $CrPS_4$ devices, as the temperature decreases, the tunneling resistance initially increases, followed by a subsequent decrease. In stark contrast to few-layer $CrI_3$ tunnel devices[39], the decrease of the tunneling resistance of the $CrPS_4$ samples at low temperatures suggests a diminished influence of the spin filtering effect, typically associated with interlayer AFM ordering in few-layer 2D magnets. Furthermore, in $CrI_3$, the pronounced increase in tunneling resistance correlates directly with the $T_N$, serving as a characteristic marker. However, this correlation does not hold for few-layer $CrPS_4$. As depicted in **Figure 1d**, the downturns in resistance occur at ~ 120 K for 2L $CrPS_4$, ~ 70 K for 3L, and ~ 55 K for 4L samples. Intriguingly, these temperatures significantly exceed the $T_N$ of 38 K for bulk $CrPS_4$, suggesting alternative mechanisms (such as metal-semiconductor/insulator transition or temperature induced distortion in crystal structure) influencing the thermal behavior of the tunneling resistance in these $CrPS_4$ thin layers.

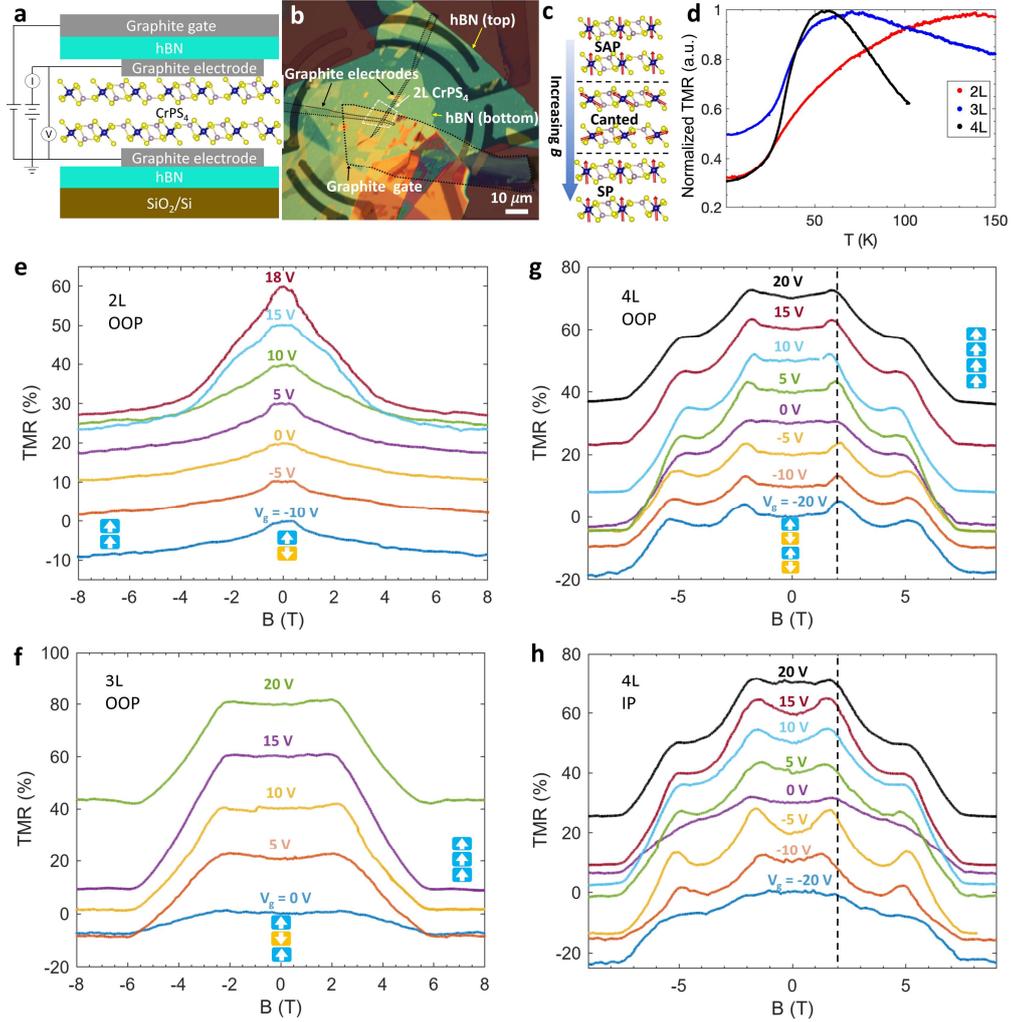

**Figure 1**. Layer- and gate-dependent magnetism in few-layer $CrPS_4$. a) Schematic of the structure of a $CrPS_4$ tunneling device with a stack of graphite/hBN/graphite /$CrPS_4$/graphite/hBN. b) Optical image of the top view of the 2L $CrPS_4$ tunneling device as illustrated in (a). c) Schematic for the evolution of spin structure under various magnetic field in $CrPS_4$. Spin-antiparallel (SAP), canted spin (Canted), and spin-parallel (SP) correspond to magnetic structure at low, intermediate, and high magnetic field, respectively. d) Temperature-dependent normalized tunneling resistance $R$ for 2L (red), 3L (blue) and 4L (black) $CrPS_4$ tunneling devices. e-g) TMR vs. $B$ at different gate voltages for 2L (e), 3L (f) and 4L (g) $CrPS_4$ tunneling devices. The bias voltages applied were -450, 700 and -600 mV for 2L, 3L and 4L $CrPS_4$, respectively. The arrows indicate the magnetization of the corresponding $CrPS_4$ layers at given magnetic fields. The dash line in (g) marks $B$ = 2 T. The applied magnetic field is along out-of-plane (OOP) direction. h) TMR as a function of gate voltage for the 4L $CrPS_4$ under an in-plane (IP) magnetic field. The dash line marks $B$ = 2 T. The measurements in (e)- (h) were performed at $T$ = 1.5 K. The TMR curves have been vertically shifted to enhance clarity.
5

We then explore the dependence of the TMR on the gate voltage for CrPS$_4$-based MTJs under both OOP and IP magnetic fields. The TMR is defined as TMR $= (R_B - R_{B=0})/R_{B=0} \times 100\%$, where $R_B$ is the tunneling resistance at a given magnetic field $B$ and $R_{B=0}$ is tunneling resistance measured at $B$ = 0 T. **Figure S2** (Supporting Information) shows the gate-dependent TMR for three CrPS$_4$-based MTJs under different OOP magnetic fields. For the 2L CrPS$_4$ device, the TMR generally becomes more negative as the gate voltage increases at different magnetic fields. For instance, at $B$ = 7 T, the TMR drops from ~ -10% to ~ -32% as the gate voltage increases from 0 V to 18 V. In the case of the 3L CrPS$_4$ device, the TMR initially decreases with the increasing gate voltage, reaching a minimum (maximum in magnitude) at around 15 V, followed by an increase upon further voltage elevation. For instance, at $B$ = 8 T, there is a significant decrease in TMR from roughly -10% to -50% as the gate voltage extends from 0 V to 15 V. For the 4L CrPS$_4$, a similar trend is observed, where the TMR first decreases with an increasing $V_g$, peaking at $V_g$ ~ 15 V, before increasing with subsequent increases in gate voltage. We note that the maximum gate voltages for the 2L, 3L, and 4L CrPS$_4$ tunneling devices, defined as the thresholds beyond which significant leakage current (less than 50 pA) is observed, are 18 V, 20 V, and 20 V, respectively. **Figure 1e-g** show the corresponding TMR as a function of the OOP magnetic field under different gate voltages for 2L, 3L and 4L CrPS$_4$, respectively. For the 2L CrPS$_4$, depicted in **Figure 1e**, the ground state at zero magnetic field is an antiferromagnetic (AFM) state denoted as ↑↓, where each arrow represents the spin orientation in the corresponding layer. As the magnetic field increases, a notable decrease in TMR is observed when the magnitude of $B$ approaches ~ 0.4 T. The gradual decrease of TMR indicates a spin-flop transition from the AFM to a canted spin state (**Figure 1c**), in contrast to a sudden change of TMR in few-layer CrI$_3$ due to the spin-flip transition[32,37]. As the magnitude of $B$ increases further, the TMR continues to decline and starts to level off (e.g., at $V_g$ = 10, 15 and 18 V), signifying a transition from the canted spin state to a fully SP state (↑↑). Importantly, we reveal a strong dependence of TMR on gate voltage. When the gate voltage is increased from -10 to 18 V, two distinct TMR characteristics emerge: 1) a significant decrease in the TMR (being more negative), and 2) the gradual decrease and onset of saturation, clearly indicating a spin canted region and the fully spin-polarized magnetic state. The 3L CrPS$_4$ device, illustrated in **Figure 1f**, exhibits a ground state characterized by interlayer AFM coupling, denoted as either ↑↓↑ or ↓↑↓. Unlike the 2L CrPS$_4$, when subjected to increasing magnetic fields, the TMR of the 3L sample initially shows a slight increase until the magnetic field reaches ~ 2.2 T. Beyond this point, TMR rapidly decreases as $B$ continues to increase, ultimately reaching saturation at ~ 5.8 T for all the gate voltages. This behavior indicates a distinct phase transition from the ↑↓↑ (or ↓↑↓) state to a fully SP (↑↑↑) state through the spin-flop process. In terms of gate voltage dependence, the TMR response of the 3L CrPS$_4$ is also notable. As the $V_g$ increases from 0 to 20 V, the difference in TMR magnitude between $B = 0$ and 9 T initially grows, peaking at $V_g = 15$ V, before subsequently diminishing. In the 4L CrPS$_4$ device, we observed a series of more intriguing features. **Figure 1g** and **1h** show the 4L sample TMR as a function of OOP and IP magnetic fields, respectively, at various gate voltages. Specifically, under most gate voltages, the TMR in 4L CrPS$_4$ exhibits a progressive increase with increasing OOP magnetic fields, culminating in a pronounced peak around 2 T. This contrasts markedly with the behavior observed in the few-layer CrI$_3$ tunneling device, where TMR, predominantly driven by the spin filtering effect, remains largely invariant until the onset of a spin-flip transition. Furthermore, the 4L CrPS$_4$ undergoes a spin-flop transition, evolving into the fully SP state (↑↑↑↑) at approximately 7.1 T. Notably, amidst this transition, we detected an additional TMR peak around 5 T for all applied gate voltages. This finding is paralleled by the observation of similar peak-like TMR features under IP magnetic fields (**Figure 1h**). Additionally, our data indicate that the gate voltage has a substantial influence on the TMR characteristics in the 4L CrPS$_4$ device, significantly affecting both the magnitude and the peak positions (black dashed lines) of the TMR curves under both OOP and IP magnetic fields.

Next, we study the intriguing peak-like features observed in the 4L CrPS$_4$ samples. Our transport results indicate these features are weak in the 2L and 3L samples. Considering the structural similarity across all devices and the occurrence of these peak-like features in 4L CrPS$_4$ under both OOP and IP magnetic fields, we can confidently discount the Shubnikov-de Haas oscillations, potentially associated with few-layer graphene electrodes, as a possible cause. Further, these peak-like features are discernible under both OOP and IP magnetic fields, with no obvious shift of the magnetic fields at the corresponding TMR peak positions for the two magnetic field orientations. Such observations lead us to speculate that these features are unlikely to be associated with the emergence of intermediate, layer-dependent spin configurations, such as that observed in few-layer CrI$_3$. This is underpinned by the fact that the magnetic structure of CrPS$_4$ predominantly exhibits an easy axis along the OOP direction, which would necessitate a substantially stronger IP magnetic field to effectuate a change in the spin state. Also, since CrPS$_4$ undergoes a spin-flop transition instead of a spin-flip transition, the characteristic peaks and valleys of TMR do not align with the spin-filtering effect. Consequently, we propose that these peak-like features could be manifestations of magnetic field-induced oscillations in TMR of few-layer CrPS$_4$. We note that these TMR oscillations have not been observed in other 2D magnets, thus far. A plausible explanation for the TMR oscillations could be the spin geometric phase mechanism between two spin current tunneling channels, where the tunneling electron spin interacts with the canted spin textures in both channels. Variations in magnetic field flux and the number of layers are likely to induce changes in the spin geometric phase[58–61], resulting in oscillations of the tunneling current. However, a comprehensive understanding of these observed oscillations necessitates further experimental and theoretical investigations. We further note that the measured TMR of few-layer CrPS$_4$ encompasses both the spin-flop transition and TMR oscillations induced by the magnetic field. The distinct variations in TMR observed in the CrPS$_4$ tunnel junction devices with different layers can be attributed to the following factors. Firstly, it is documented that 2D magnets often exhibit layer-dependent coercivity for TMR transitions particularly in the few-layer region, as evidenced in other 2D magnetic systems like CrI$_3$. Secondly, our results indicate that TMR oscillations are more pronounced in thicker layers.

**2.2. Electrically Tunable Tunneling Anisotropic Magnetoresistance**

It is well-accepted that anisotropic magnetoresistance (AMR), which describes how the magnetoresistance changes with magnetic field direction in a ferromagnetic conducting film, leads to the relation: $\rho(\theta) = \rho_\perp + (\rho_\parallel - \rho_\perp)cos^2\theta$, where $\rho$ is the resistivity of the film, $\theta$ is the angle between the direction of magnetization and current, and $\rho_\parallel$ and $\rho_\perp$ are resistivities at $\theta = 0°$ and $\theta = 90°$, respectively. The origin of AMR can be traced back to spin-orbital coupling. When considering the tunneling transport, a more complex bias-dependent TAMR was observed in conventional MTJs[62] such as CoFe/MgO/CoFe and CoFe/Al$_2$O$_3$/CoFe. However, TAMR in 2D magnet-based MTJs remains elusive. In the few-layer CrPS$_4$-based MTJs, we investigate the influence of three key parameters on the TMR: 1) the bias voltage and 2) gate voltage applied to the junction as well as 3) the angle between the external magnetic field and the IP direction of the few-layer CrPS$_4$.

We find that both the polarity and magnitude of the TAMR in these MTJs can be extensively modulated through electrical methods. Here, we define the angle $\theta_B$ as the angle between the $B$ field and the sample plane, as shown in **Figure 2a**. The TAMR is calculated using the formula TAMR $= (R_{\theta_B} - R_{\theta_B=0°}) \times 100\%/R_{\theta_B=0°}$, where $R_{\theta_B}$ is the tunneling resistance at a given magnetic field and $R_{\theta_B=0°}$ represents the tunneling resistance at $\theta_B = 0°$. **Figure 2b** shows a colormap depicting the TAMR of the 2L CrPS$_4$ MTJ as a function of $\theta_B$ under various bias voltages at $B = 7$ T and $T = 1.5$ K. We note that the TAMR is ~ 0% at both $\theta_B = 0°$ and 180° and the corresponding magnetic states of the few-layer CrPS$_4$ are in the fully SP state. **Figure 2c** exhibits the line cuts of TAMRs for selected biases, as derived from Figure 2b. We find that the TAMR has a two-fold rotational symmetry regardless of the applied bias, as shown Figure 2b and 2c. This behavior can be primarily attributed to the even symmetry typically observed in the TMR of few-layer 2D magnets relative to the applied magnetic field. Remarkably, as the bias voltage varies from -620 to -320 mV, we see that the TAMR of 2L CrPS$_4$ undergoes a change in magnitude and a sign reversal. For instance, at a bias voltage (e.g., -320 mV), the TAMR is negative, displaying peaks at $\theta_B = 0°$ and 180°, while the valleys occur at $\theta_B = 90°$ and 270°. Conversely, at a more negative bias (- 620 mV), the TAMR becomes positive, with valleys at $\theta_B = 0°$ and 180°, and peaks at $\theta_B = 90°$ and 270°. Consequently, the TAMR at $\theta_B = 90°$ increases from -2% to 4% when the bias voltage changes from -320 mV to -620 mV. We further demonstrate that the bias-voltage tunable TAMR can also be realized in the 4L CrPS$_4$.

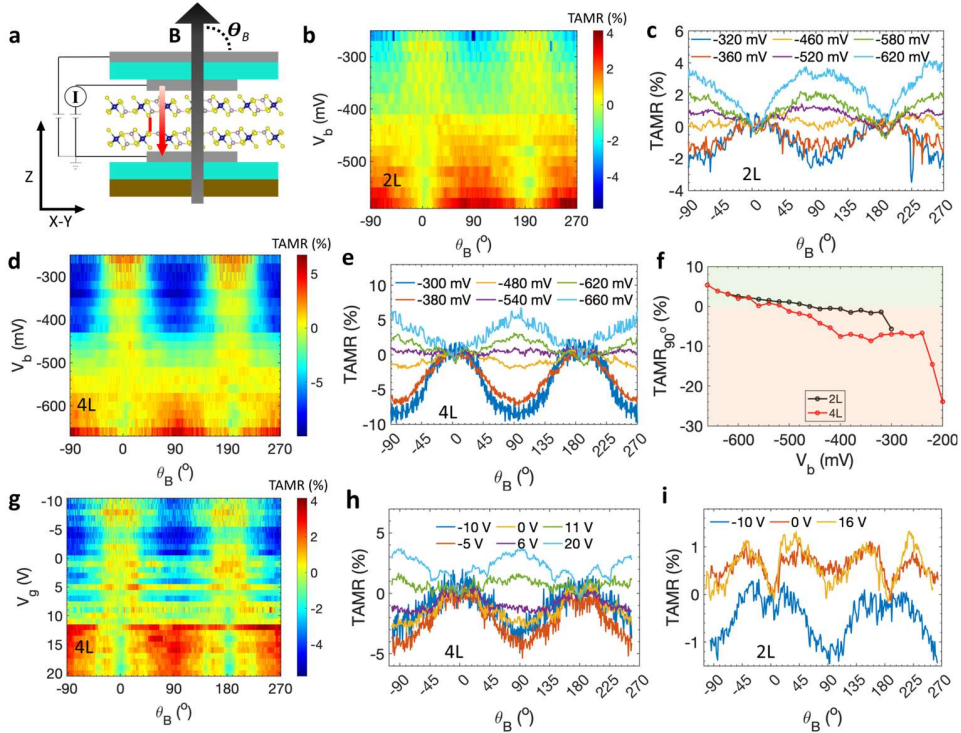

**Figure 2.** Bias- and gate-dependent TAMR in few-layer CrPS$_4$. a) Schematic for the definition of angle of $B$ field $\theta_B$. The black arrow indicates the $B$ field direction. The red arrow indicates the current direction. b) TAMR as a function of $\theta_B$ at different bias voltages for the 2L CrPS$_4$ tunneling device. The measurements were performed at $T = 1.5$ K. The applied $B$ field is 7 T. c) TAMR as a function of $\theta_B$ at selected bias voltage $V_b$ taken from (b). d) TAMR as a function of $\theta_B$ at different bias voltages for the 4L CrPS$_4$ tunneling device. The measurements were performed at $T = 1.5$ K. The applied $B$ field was 9 T. e) TAMR as a function of $\theta_B$ at selected bias voltage $V_b$'s taken from (d). f) TAMR at $\theta_B = 90°$ as a function of bias voltage for the 2L and 4L CrPS$_4$ tunneling devices. The data are taken from (b) and (d). g) TAMR as a function of $\theta_B$ at different gate voltages $V_g$ for the 4L CrPS$_4$ tunneling device. The measurements were performed at $T = 1.5$ K. The applied $B$ field was 9 T. h) TAMR as a function of $\theta_B$ at selected gate voltage $V_g$'s taken from (g). i) TAMR as a function of $\theta_B$ at selected gate voltage $V_g$'s for the 2L CrPS$_4$ tunneling device at $T = 1.5$ K and $B = 7$ T.

**Figure 2d** shows the corresponding colormap of the TAMR vs. $\theta_B$ under different bias voltages measured at $T = 1.5$ K and $B = 9$ T. **Figure 2e** shows the line cuts of TAMRs taken from Figure 2d for selected biases. **Figure 2f** displays the TAMR for both the 2L (black dots) and 4L (red dots) CrPS$_4$ tunneling devices, measured at $\theta_B = 90°$, and plotted as a function of bias voltage. We see that the TAMR decreases as the magnitude of the bias voltages reduces. The critical bias voltages for sign reversal are about -420 mV and -520 mV for 2L and 4L CrPS$_4$ MTJs, respectively. We note that a similar bias voltage-tunable TAMR behavior can also be observed in 3L CrPS$_4$ (see **Figure S3** in Supporting Information) and an extended positive bias range in the 4L CrPS$_4$ (see **Figure S4** in Supporting Information), suggesting that this phenomenon is a universal characteristic of few-layer CrPS$_4$-based MTJs.

We further explore the dependences of both the polarity and magnitude of TAMR on the applied gate voltages. **Figure 2g** is the colormap that depicts the TAMR of the 4L CrPS$_4$ device as a function of $\theta_B$ under various gate voltages at a fixed $B$ field of 9 T and $T = 1.5$ K. The line cuts corresponding to selected gate voltages are shown in **Figure 2h**. We observe that as the gate voltage changes from negative to positive, the sign of the TAMR transitions from negative to positive, exhibiting a reversal behavior similar to that observed under changes in the bias voltage. For the 2L CrPS$_4$ tunneling device, we see that the gate voltage can also tune the magnitude and the sign of TAMR, as shown in **Figure 2i**. However, the corresponding gate voltage effect on the TAMR in 2L CrPS$_4$ seems weaker than that of the 4L sample. In general, we conclude that gate and bias voltages play a similar role in controlling the polarity and magnitude of TAMR in few-layer $CrPS_4$-based MTJs. We also notice that, compared to the 4L $CrPS_4$, the distorted gate voltage dependence of TAMR in the 2L $CrPS_4$ device (**Figure 2i**) suggests a possible doping-induced alteration in the crystal structure and hence the magnetic ground state. In addition, upon comparison with **Figure 1**, it becomes evident that the data presented in **Figure 2** displays higher noise levels and noticeable sharp peaks. This discrepancy can be attributed to the significantly smaller magnitude of the TAMR observed in **Figure 2**, leading to the amplified noise levels in the curves.

The electrically tunable TAMR in the 4L $CrPS_4$ has also been observed in the SP phase region and the magnetic states (such as a canted spin state and an AFM state) under considerably lower magnetic fields. **Figure 3a** shows the corresponding colormap of TAMR versus $\theta_B$ under different bias voltages measured at $B = 6$ T, a condition that places the 4L $CrPS_4$ in a canted spin phase (as depicted in Figure 1g). The behavior of TAMR *vs.* $\theta_B$ under selected bias voltages from **Figure 3a** is further detailed in **Figure 3b**. We see that the TAMR at $\theta_B$ = -90°, 90° and 270° switches from negative to positive as the bias voltage becomes more negative. This bias-dependent sign change is similarly observed under lower background magnetic fields of 4.5 T (where the 4L CrPS$_4$ is in a canted spin state) and 1 T (where the 4L CrPS$_4$ is in the AFM state). In the 2L CrPS$_4$-based MTJ, a comparable bias voltage-dependent TAMR is also detected in the canted spin phases ($B$ = 5 T and 1 T, as detailed in **Figure S5** in Supporting Information). Additionally, the gate-dependence of TAMR in the AFM state (at $B$ = 1 T) is demonstrated in **Figure S6** (Supporting Information). These experimental findings collectively reinforce the notion that the electrically tunable nature of TAMR is a universal characteristic across few-layer CrPS$_4$ devices with different spin configurations. We note that these interesting features have not been reported thus far.

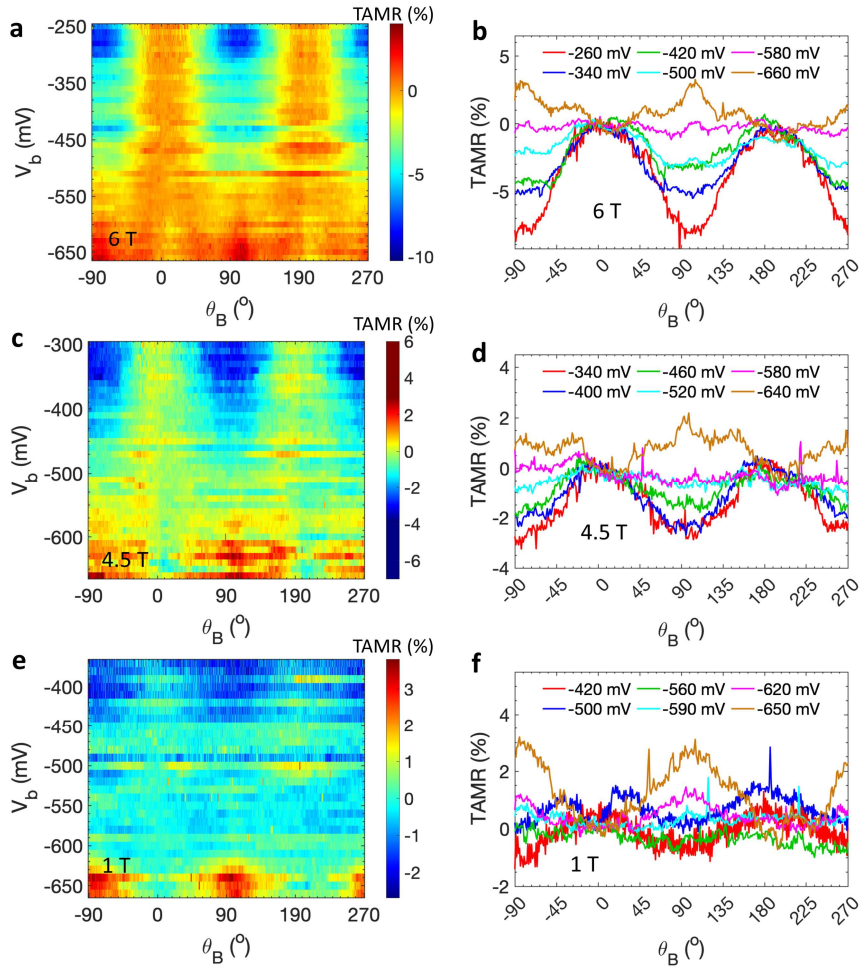

**Figure 3.** Bias-dependent polarity of TAMR in the 4L CrPS$_4$ tunneling device at $B$ = 6, 4.5 and 1 T. a) TAMR as a function of $\theta_B$ at different bias voltages for the 4L CrPS$_4$ tunneling device. The measurements were performed at $T$ = 1.5 K. The applied $B$ field was 6 T. b) TAMR as a function of $\theta_B$ at selected bias voltage $V_b$'s taken from (a). c) The same as (a) but the applied $B$ field is 4.5 T. d) TAMR as a function of $\theta_B$ at selected bias voltage $V_b$'s taken from (c). e) The same as (a) but the applied $B$ field is 1 T. f) TAMR as a function of $\theta_B$ at selected bias voltage $V_b$'s taken from (e).

**2.3. First-principles Calculations to Understand the Electrically Tunable TAMR in CrPS$_4$**

To gain an insight into the origin of electrically tunable and polarity reversal of the TAMR, we employ first-principles calculations to elucidate the tunneling transport in the OOP and IP directions of a bilayer CrPS$_4$-based tunneling junctions with top and bottom graphene electrodes. To simulate systems with bias voltages, in the calculations, we added an electric field ranging from 0.02 to 0.21 V/Å along the z-axis (OOP), corresponding to the interlayer bias voltage from 0.043 V to 0.365 V in the 2L CrPS$_4$. Based on the non-equilibrium Green's function (NEGF) method[63], the corresponding tunneling current $I$ with bias voltage $V$ is calculated along the OOP direction. The Wannier function (WF)-based tight-binding Hamiltonian of the 2L CrPS$_4$ from first-principles calculations is used as the central part. To simplify the device structure, the single $s$ orbital non-magnetic atomic chain is treated as the lead, which connects to the $p_z$ orbitals of the top and bottom S atoms. No barrier between the leads and the 2L CrPS$_4$ is assumed. As shown in **Figure 4a**, with p-type doping, the chemical potentials for the left ($\mu_l$) and right ($\mu_r$) leads are set to the valence band maximum (VBM) of the upper and lower layers, respectively, while with n-type doping, those are set to the conduction band minimum (CBM) of both layers. The transmission rate $T(E)$ as a function of $E$, the kinetic energy of the injection electrons from leads, is obtained with a $96 \times 96$ k-points mesh in the Brillouin zone. So that the tunneling current $I$ is given by

$$I = -\frac{e}{h}\int_E dE\, T(E)[f(E-\mu_l) - f(E-\mu_r)]$$

where $e$ is the electron charge, $h$ is Planck's constant, and $f(E)$ is the Fermi-Dirac distribution function with the temperature $k_B T = 1.0$ meV. Here, $V = -1/e\,(\mu_l - \mu_r)$ is the bias voltage. $I-V$ relationship is obtained for two ferromagnetic spin states: one with (001) spin direction and the other with (100) spin direction, corresponding to the spin-polarized states under the OOP and IP magnetic fields of about 7 T, respectively. **Figure 4b** shows the difference rate of tunneling current, $\frac{I(001)-I(100)}{I(001)} \times 100\%$, equivalent to the tunneling anisotropic magnetoconductance, as a function of bias voltage for both n-type and p-type doping situations. In the p-type doping region, a transition from a positive to a negative value appears at about $V = 0.25\text{V}$. In this model, positive and negative bias voltages have the symmetric $I-V$ relation so that at about $V = -0.25\text{V}$, a similar sign reversal is expected, consistent with the experimental results (also see Supplementary Note 1). The resistance with OOP magnetization $R_\perp$ is lower than that with in-plane magnetization $R_\parallel$, while $R_\perp > R_\parallel$ with $|V| > 0.25\text{V}$. The 1% difference of the magnitude is consistent with the experimental observation between $\theta_B = 0°$ and $90°$, and is mainly from the contribution of band shift of all bands between $\mu_l$ and $\mu_r$ due to anisotropic spin directions (See **Figure S7** in Supporting Information).

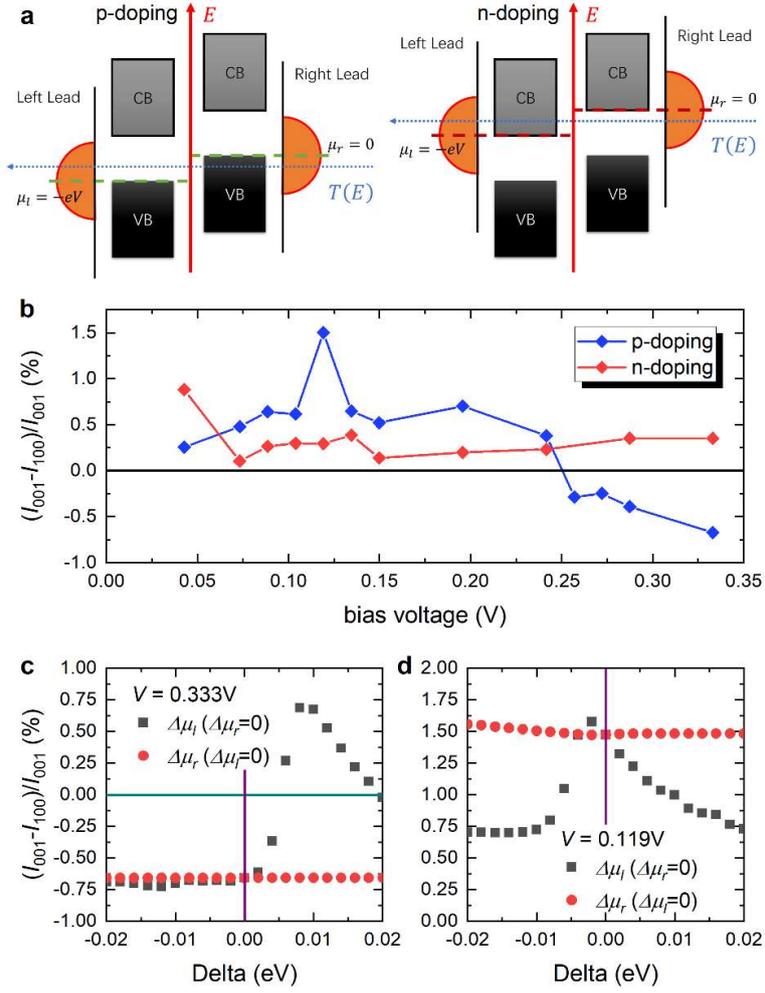

**Figure 4.** First-principles calculation for tunneling current along OOP (001) and IP (001) directions in bilayer CrPS$_4$. a) The tunneling model of the bilayer CrPS$_4$ under bias $V$, left and right panels correspond to p-doping and n-doping, respectively. CB and VB correspond to conduction bands and valence bands, respectively. b) The difference rate of tunnelling current $\frac{[I(001)-I(100)]}{I(001)} \times 100\%$ as a function of bias voltage. With p-doping and $V = 0.333\text{V}$ (c) and $0.119\text{V}$ (d), the rate $\frac{[I(001)-I(100)]}{I(001)} \times 100\%$ as a function of a small shift $\Delta$ of $\mu_l$ ($\Delta\mu_r = 0$) and $\mu_r$ ($\Delta\mu_l = 0$).

Both bias and gate voltages not only control the bias between two CrPS$_4$ layers but also potentially adjust the chemical potential for each layer. An energy offset should actually appear between the chemical potential and VBM (p-type doping)/CBM (n-type doping), leading to an asymmetric $I - V$ relationship. To simulate this asymmetric behavior, we made small shifts $\Delta$s of $\mu_l$ and $\mu_r$, namely $\Delta\mu_l$ and $\Delta\mu_r$ respectively. Under the p-type doping environment, **Figure 4c** and **4d** show the rate $\frac{I(001)-I(100)}{I(001)} \times 100\%$ as a function of $\Delta\mu_l$ and $\Delta\mu_r$ under = 0.333V and 0.119V, respectively. While only tiny change happens with finite $\Delta\mu_r$, positive and negative $\Delta\mu_l$ lead to distinct behaviors, which eventually cause the asymmetric $I - V$ curve.

Another possible mechanism behind the electrically tunable TAMR is involving a switch in the easy axis from OOP to IP orientation due to the applied bias or gate voltage [64]. To investigate this possibility, we conducted a thorough analysis of our results under both bias and gate voltage conditions. In the case of gate voltage, as depicted in **Figures 2h** and **S6** for a 4L sample, we compared the changes in TAMR magnitude with varying gate voltages and magnetic fields (9 T and 1 T). Notably, at $\theta_B$ = 90°, the TAMR magnitude shifts from 4% to -5% at 9 T and from 3% to -3% at 1 T. These findings suggest that while doping-induced variation in the easy axis orientation cannot be discounted, their impact on the observed TAMR polarity switch appears limited. This assertion is supported by the fact that, if doping were the primary driver, a more significant effect would be expected at lower magnetic fields (1 T) compared to higher fields (9 T). Similarly, under bias conditions (**Figure 3**), we see a decrease in TAMR magnitude as the magnetic field reduces. These results suggest that while bias or gate voltage-induced changes in the easy axis orientation may contribute, they are not the dominant factor driving the observed TAMR polarity switching.

## 3. Conclusion

In conclusion, we have investigated the layer-dependent behavior of TMR in few-layer CrPS$_4$-based MTJs. This investigation has not only illuminated the dependence of TMR on gate voltage but has also uncovered the anomalous TMR oscillations under both OOP and IP magnetic fields, potentially heralding the presence of a spin geometric phase. A cornerstone of our study is the discovery of a highly tunable TAMR in the few-layer CrPS$_4$ through both bias and gate voltages. This adaptability of TAMR, discernible across diverse states of CrPS$_4$, including AFM, canted spin, and SP phases under various magnetic fields, marks a significant advancement. Theoretical support for these experimental observations was provided through first-principles calculations, which helped delineate the differences in electrical transport properties between IP and OOP directions in a 2L CrPS$_4$ system. Our findings not only advance the understanding of 2D magnetic materials but also uncover the novel quantum states, opening new opportunities in the realm of materials science and spintronics.

## 4. Methods

### 4.1. Synthesis of CrPS$_4$ crystals and device fabrication

Single crystals of CrPS$_4$ were grown by chemical vapor transport using iodine as the transport media. While under an atmosphere of pure argon, 0.52 g of metallic chromium (BTC 99.99%), 0.31 g of elemental phosphorus (Millipore-Sigma 99.99%), 1.30 g of sulfur powder (Alfa-Aesar 99.5%) and 0.025 g of iodine (Thermo-Fisher 99.99) were placed in a fused silica tube (0.9 cm inner diameter × 20 cm length). The tube was then cooled to 77 K, evacuated to 50 mTorr, and then sealed to an over-length of 15 cm. The tube contents were pre-reacted by heating at 25 K/min to 525 K, and maintaining that temperature for 20 hours after which they were cooled to room temperature at 5 K/min. After vigorous shaking of the tube to re-mix the reagents, the tube was placed in a two-zone furnace and heated at 5 K/min to 900 °C at the charge zone and to 875 K at the growth zone. The tube was held at these conditions for 175 hours, after which it was cooled at 25 K/min to room temperature. The tube was then opened under an argon atmosphere, the crystals mechanically extracted, then sealed in glass scintillation vials for further use.

The few-layer CrPS$_4$ tunneling junction was fabricated by a layer-by-layer dry transfer method[65,66]. Atomically thin CrPS$_4$, hBN and graphite were mechanically exfoliated from their bulk crystals onto the SiO$_2$(200 nm)/Si substrates. The sample thickness was determined by an atomic force microscope. For a monolayer CrPS$_4$, the thickness is of ~ 0.7 nm. The quality of our exfoliated CrPS$_4$ flakes was further characterized by Raman spectroscopy (See **Figure S8** in Supporting Information). The stack of hBN/graphite/CrPS$_4$/graphite/hBN was picked up one by one using a polydimethylsiloxane stamp with a polyvinyl alcohol (PVA) layer on the top. The entire stack was then released onto a SiO$_2$/Si substrate with prefabricated Pt/Ti (30nm/5nm) electrodes, which were prepared by a standard nanofabrication procedure. After dissolving the PVA layer in deionized water, a graphite flake was finally transferred onto the stack to serve as a top gate. Here, the hBN flakes serve as protection and dielectric layers, and graphite flakes serve as either bias or gate electrodes. To avoid any possible degradation of the thin $CrPS_4$ layers, the exfoliation and the transfer processes were performed in an argon-filled glove box with $H_2O$ and $O_2$ concentrations of < 0.1 ppm.

### 4.2. Electrical and magnetotransport measurements

The low temperature electrical and magnetotransport measurements were carried out inside a closed cycle $^4$He cryostat (Oxford TeslatronPT with a base temperature of 1.5 K). The angle between the sample and magnetic field was controlled by rotating the sample via a self-written program. The DC electrical transport measurements were performed with a Keithley 2400 Sourcemeter. We note that all the results for 2L (or 3L and 4L) (or "for a specific layer) $CrPS_4$ tunneling device in this work were obtained from the same device.

### 4.3. Raman spectroscopy measurements

The Raman spectra were acquired through a Horiba confocal Raman microscope with a 532 nm laser excitation. A 2400 grooves/mm grating was used to achieve a spectral resolution of below 1.4 cm$^{-1}$. The Raman spectra taken at different temperatures were conducted in a cooling stage down to liquid nitrogen temperature with an optical window (INSTEC, INC, model HCP421V-PMH).

### 4.4. First-principles calculations

Our first-principles calculations are performed to calculate electronic and magnetic properties of bilayer $CrPS_4$ by using the projector-augmented wave pseudopotential [67] implemented in the VASP package[68,69]. Generalized gradient approximation in Perdew–Burke–Ernzerhofer formation[70] is employed as the exchange–correlation potential, and the Hubbard $U$ method [71] is introduced to treat localized 3d orbitals of Cr atoms, using $U = 2.5$ eV as previously tested. An energy cutoff of 500 eV is used for the plane-wave expansion throughout the calculations. The Γ-centered 2D k-points mesh of 5×7 is sampled in the Brillouin zone. A vacuum region of 15 Å is chosen to prevent artificial interactions between neighboring sheets along the z direction. The interlayer interactions in the bilayer $CrPS_4$ are considered by adopting the DFT-D3 method [72] to describe long-ranged van der Waals interactions. The structures are fully relaxed until the force on each atom is smaller than 0.01 eV/Å, and the total energy convergence criterion is set as $10^{-7}$ eV. Spin-orbit coupling is included in self-consistent electronic calculations.

A unitary transformation of Bloch waves was performed to construct the tight-binding Hamiltonian in a WF basis implemented in the WANNIER90 package[73]. A WF-based

Hamiltonian has exactly the same eigenvalues as those obtained by first-principles calculations among all occupied bands and the bands below 0.5eV to CBM.

**Acknowledgements**

J.T. also acknowledges the financial support of the U.S. National Science Foundation (NSF) grant 2228841 for data analysis, H. H. and J. Y. acknowledge the financial support of the National Natural Science Foundation of China (Grant Number 12274309), K.W. and T.T. acknowledge support from the JSPS KAKENHI (Grant Numbers 20H00354 and 23H02052) and World Premier International Research Center Initiative (WPI), MEXT, Japan.